\pdfoutput=1
\documentclass[prd,aps,nofootinbib,twocolumn,superscriptaddress,preprintnumbers,balancelastpage,longbibliography]{revtex4-1}

\usepackage{graphicx}	
\usepackage{amsmath}	
\usepackage{graphics}
\usepackage{afterpage}
\usepackage{float}
\usepackage{subfigure}
\usepackage{rotating}
\usepackage{multirow}
\usepackage{tabularx}
\usepackage{booktabs}
\usepackage{fancyhdr}
\usepackage{hyperref}
\usepackage[utf8]{inputenc}
\usepackage{theorem}
\usepackage{moreverb}
\usepackage{euscript}
\usepackage{psfrag}
\usepackage{slashed}
\usepackage{mathtools}
\usepackage{makecell}
\usepackage[flushleft]{threeparttable}
\usepackage{adjustbox}

\usepackage{dcolumn}
\usepackage{bm}
\usepackage[mathlines]{lineno}
\usepackage{lettrine}
\usepackage[dvipsnames]{xcolor}
\usepackage{url}

\usepackage{graphicx}
\usepackage{bm}
\usepackage{times}
\usepackage{slashed}
\usepackage{color}
\usepackage{aas_macros}
\usepackage{siunitx}
\usepackage{slashed}
\usepackage{lipsum}
\usepackage{subfigure}
\usepackage{multirow}
\usepackage{amsmath}
\usepackage{xspace}
\usepackage{array} % for defining a new column type
\usepackage{varwidth} %for the varwidth minipage environment

\newcommand{\be}{\begin{equation}}
\newcommand{\ee}{\end{equation}}
\newcommand{\bea}{\begin{eqnarray}}
\newcommand{\eea}{\end{eqnarray}}

\newcommand{\sn}{SN\xspace}
\newcommand{\sne}{SNe\xspace}

\usepackage{listings}
\usepackage{color,xcolor}

\newcolumntype{C}[1]{>{\centering\let\newline\\\arraybackslash\hspace{0pt}}m{#1}}

\hypersetup{
     colorlinks   = true,
     citecolor    = blue,
     urlcolor     = blue,
     linkcolor    = blue
}

\begin{document}

\preprint{LTH-1321}

\title{
%Distant Beacons of Dark Matter Impacts: \\ Triggered  White Dwarf Explosions in Dwarf Spheroidal Galaxies \\
White Dwarfs in Dwarf Spheroidal Galaxies:\\
A New Class of Compact-Dark-Matter Detectors}
\author{Juri Smirnov}
\thanks{{\scriptsize Email}: \href{mailto:juri.smirnov@liverpool.ac.uk}{juri.smirnov@liverpool.ac.uk}; {\scriptsize ORCID}: \href{http://orcid.org/0000-0002-3082-0929}{0000-0002-3082-0929}}
\affiliation{Department of Mathematical Sciences, University of Liverpool,
Liverpool, L69 7ZL, United Kingdom}
\affiliation{The Oskar Klein Centre, Department of Physics, Stockholm University, AlbaNova, SE-10691 Stockholm, Sweden}
%\affiliation{Stockholm University and The Oskar Klein Centre for Cosmoparticle Physics,  Alba Nova, 10691 Stockholm, Sweden}
\author{Ariel Goobar}
\email{ariel@fysik.su.se, ORCID: orcid.org/0000-0002-4436-0820}
\affiliation{The Oskar Klein Centre, Department of Physics, Stockholm University, AlbaNova, SE-10691 Stockholm, Sweden}
%\affiliation{Stockholm University and The Oskar Klein Centre for Cosmoparticle Physics,  Alba Nova, 10691 Stockholm, Sweden}
\author{Tim Linden}
\email{linden@fysik.su.se, ORCID: orcid.org/0000-0001-9888-0971}
%\affiliation{Stockholm University and The Oskar Klein Centre for Cosmoparticle Physics,  Alba Nova, 10691 Stockholm, Sweden}
\affiliation{The Oskar Klein Centre, Department of Physics, Stockholm University, AlbaNova, SE-10691 Stockholm, Sweden}
\author{Edvard Mörtsell}
\email{edvard@fysik.su.se, ORCID: orcid.org/0000-0002-4436-0820}
\affiliation{The Oskar Klein Centre, Department of Physics, Stockholm University, AlbaNova, SE-10691 Stockholm, Sweden}
%\affiliation{Stockholm University and The Oskar Klein Centre for Cosmoparticle Physics,  Alba Nova, 10691 Stockholm, Sweden}

\begin{abstract}
Recent surveys have discovered a population of faint supernovae, known as Ca-rich gap transients, inferred to originate from explosive ignitions of white dwarfs. In addition to their unique spectra and luminosities, these supernovae have an unusual spatial distribution and are predominantly found at large distances from their presumed host galaxies. We show that the locations of Ca-rich gap transients are well matched to the distribution of dwarf spheroidal galaxies surrounding large galaxies, in accordance with a scenario where dark matter interactions induce thermonuclear explosions
among low-mass white dwarfs that may be otherwise difficult to ignite with standard stellar or binary evolution mechanisms. A plausible candidate to explain the observed event rate are primordial black holes with masses above $10^{21}$ grams. 
\end{abstract}

\maketitle

Recent surveys have uncovered a new population of supernovae (SN) with peculiar properties~\cite{2010Natur.465..322P,2017ApJ...836...60L}, called Ca-Rich Gap Transients. Compared to standard Type Ia~SN, which trace stellar density, Ca-rich events are located at a much larger offset from the center of their host galaxies. Additionally, spectral observations seem to indicate that they originate from white dwarfs with masses well below the Chandrasekhar limit, at around $\sim$~0.6 $M_{\odot}$~\cite{2015A&A...573A..57M}. Finally, these events predominantly occur in old systems, such as elliptical galaxies. 

White dwarfs (WDs) are essentially nuclear bombs in space --- waiting to be ignited by sufficient energy injection. We propose that Ca-Rich transients are naturally explained in a scenario where the WDs can be ignited by interactions with compact dark matter (DM) relics~\cite{Fayet:2006sa, Bramante:2015cua, Acevedo:2019gre, Chan:2020ijt,Witten:1984rs, Graham:2018efk, Acevedo:2021kly,Carr:2009jm,Carr:2016drx, Graham:2015apa, Janish:2019nkk, Steigerwald:2021bro}. 

This scenario naturally explains the observed spatial distribution of Ca-rich transients for two reasons. First, the abundance of low mass white dwarfs is about an order of magnitude larger in systems that are older than eight Gyrs, compared to systems with ages below three Gyrs. Second, the event rate is dominated by ignition in dwarf spheroidal galaxies (dSphs) surrounding the host galaxy, which leads to a clustering at large radii. %We discuss in detail which conditions make the satellites ideal for dark matter-white dwarf ignition events.

While many DM models are possible, if a fraction of the DM is composed of primordial black holes (PBH), then the observed Ca-Rich transient event rate points towards the existence of asteroid mass PBHs that lie within reach of current or near-future micro-lensing surveys~\cite{Niikura:2017zjd,Smyth:2019whb}. Furthermore, deep follow-up observations of the Ca-Rich event locations are expected to reveal their correlation with dSphs surrounding distant galaxies. Thus, we find that supernovae triggered by dark matter impacts could serve as beacons for a new type of dark matter search.

\begin{figure}[t!]
\vspace{-4mm}
    \centering
    \includegraphics[width=0.9\columnwidth]{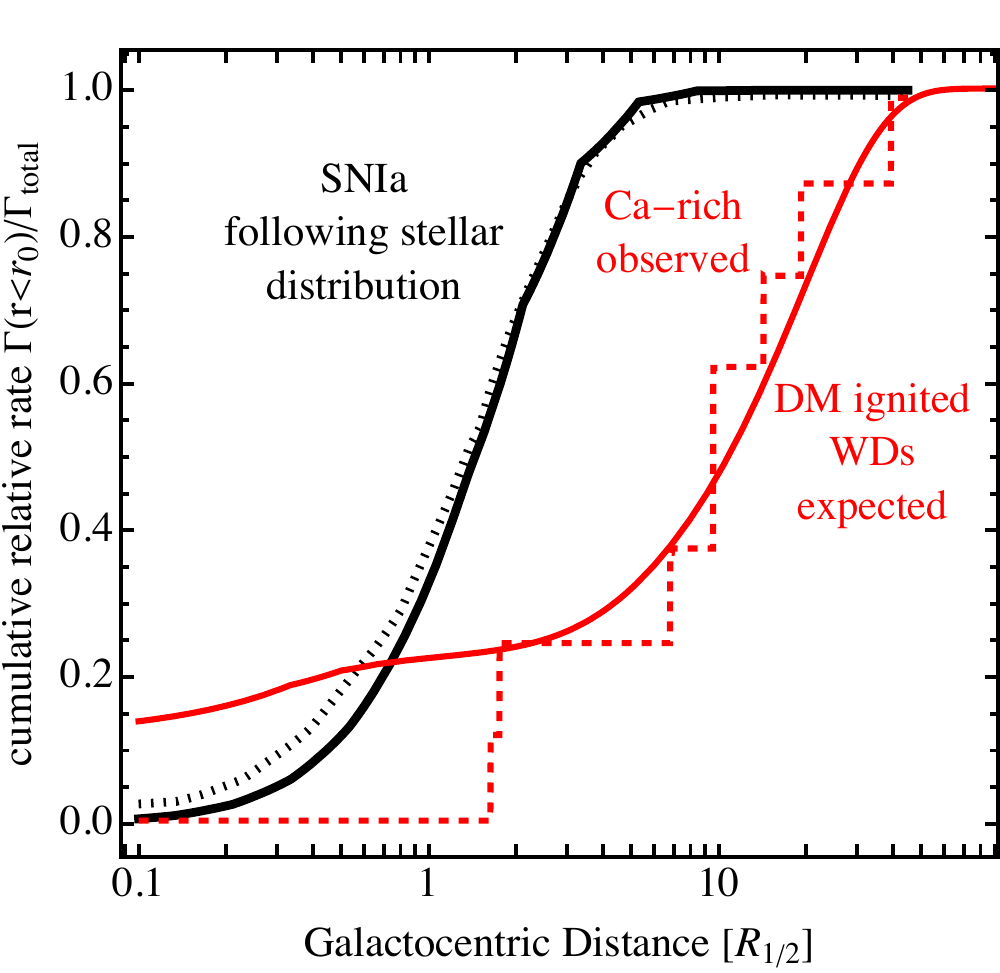}
    \caption{Radial distributions of standard Type Ia SNe (black dashed) and Ca-rich gap transients (red dashed) within their host galaxies, as determined by Ref.~\cite{2017ApJ...836...60L}. Results are compared to the spatial distribution of galactic stars (black solid), and the expected spatial distribution of DM/WD interactions (red solid).
    %, including contributions from dwarf spheroidal galaxies (blue solid).
    Ca-rich transients closely fit the morphology expected from DM/WD interactions.}
    \vspace{-4mm}
    \label{fig:datacompare}
\end{figure}

In Fig.~\ref{fig:datacompare} we show the cumulative Type Ia \sn (SNIa) event rate normalized to the half light radii of the host galaxies, as determined by Ref.~\cite{2017ApJ...836...60L}. For standard SNIa events, the observed locations closely follow the stellar distribution, matching expectations for events driven by binary interactions. Ca-rich transients, on the other hand, have a significantly extended spatial profile that does not match the stellar distribution. Intriguingly, we find that the galactocentric distance profile of these \sne is well-fit by the expected distribution of dark matter/white dwarf interactions, including a significant contribution from the dwarf spheroidal galaxies that are expected to surround the SN host galaxies. This close spatial match, taken in addition to the unique chemical properties and atypical progenitors of Ca-rich transients, motivates an interpretation where these events may be driven by dark matter/compact object interactions. \\

\noindent {\bf \emph{Ca-Rich Gap Transients.}---}
Ca-rich transients are identified based on their spectral and morphological properties~\cite{2010Natur.465..322P}. They are predominantly found in the outskirts of elliptical galaxies, with offsets from the presumed host that can reach $\sim$100 kpc \cite{2017ApJ...836...60L}, in contrast to the radial stellar density profiles of galaxies as well as both Type Ia and core-collapse SN. Ca-rich transients also differ from other SN classes in that they evolve faster to the nebular emission phase, where they also become "Ca-rich", as defined by the integrated flux ratio in [Ca II] ($\lambda \lambda 7291, 7324$) over [O I] ($\lambda \lambda 6300, 6364$) \cite{2012ApJ...755..161K}.

%Ca-rich transients are a newly identified class of faint \sn whose origin is unknown \cite{2010Natur.465..322P}. What makes them particularly intriguing is that they are predominantly found in the outskirts of elliptical galaxies. The physical offset from the presumed host can reach $\sim$100 kpc \cite{2017ApJ...836...60L}, in stark contrast to the radial stellar density profiles of galaxies and the locations of all other \sn types: Type Ia and core-collapse \sne. Furthermore, they also differ from other SN classes in that they evolve faster to the nebular emission phase, where they also become "Ca-rich", as defined by the integrated flux ratio in [Ca II] ($\lambda \lambda 7291, 7324$) over [O I] ($\lambda \lambda 6300, 6364$) \cite{2012ApJ...755..161K}. 

Because they are found in old environments and their spectra lack hydrogen features, Ca-rich transients are hypothesized to stem from low-mass white dwarfs~\cite{2015A&A...573A..57M}. However, compared to "normal" Type Ia \sne, which are typically modeled as the explosion of a white dwarf in a binary system, Ca-rich transients have much lower peak brightness, occupying a ``gap" between novae and supernovae. Additionally, their narrow lightcurves indicate lower-mass ($\sim$0.6~M$_\odot$, or even lower~\cite{Sell:2015rfa}) white dwarf progenitors, compared to standard binary models. Because of their faintness, current \sn surveys have only found these transients in a relatively small volume compared to other \sne. Hence, the Ca-rich transient sample is small. After correcting for selection effects, their rate is found to be $0.13^{+0.06}_{-0.04} \times$ the Type Ia SN rate, or $3.19^{+ 1.45}_{-0.96} \times 10^{-6}$ Mpc$^{-3}$yr$^{-1}$ \cite{2020ApJ...905...58D}. 

Given the difficulty in modeling these \sne as the result of stellar explosions in binary systems \cite[see e.g.,][for recent ideas]{2021ApJ...906...65P, 2022arXiv220713110Z}, we now consider the possibility that Ca-rich transients are the result of the explosion of a {\em single} white dwarf (WD), where the thermonuclear run-away is triggered by the passage of DM. To explore this hypothesis, we investigate the expected galactocentric distances of such explosions, taking into account the spatial distribution of WDs and dark matter, and their relative velocities. \\

%Despite a lower event statistic so far, the spatial distribution of those events could be a crucial signature that leads toward decisive tests of dark matter interactions. 

%The model predictions for SNIa events, which are expected to closely follow the stellar population, black solid line, and the observed SNIa events, as reported in Ref.~\cite{2017ApJ...836...60L}, black, dashed line, are in excellent agreement. The red solid line, which shows the expected rate of triggered WD explosions, dominated by events in the small satellite galaxies surrounding the host, describes the spatial distribution of Calcium-rich gap transients with surprising accuracy, dashed red line. Despite a lower event statistic so far, the spatial distribution of those events could be a crucial signature that leads toward decisive tests of dark matter interactions. 

%\newpage

%Search for non-Standard Dark Matter interactions
\noindent {\bf \emph{Dark Matter Induced White Dwarf Ignition.}---} The high density of compact objects makes them a natural target for searches of rare DM interactions. The possibility that DM serves as a catalyst that drives SNe provides a highly luminous signature that can be examined at extragalactic distances. Models with particle~\cite{Fayet:2006sa, Bramante:2015cua, Acevedo:2019gre, Chan:2020ijt}, composite~\cite{Witten:1984rs, Graham:2018efk, Acevedo:2021kly}, compact object~\cite{Asadi:2021pwo,Asadi:2021yml,Gross:2021qgx}, or primordial black hole (PBH)~\cite{Carr:2009jm,Carr:2016drx, Graham:2015apa, Janish:2019nkk, Steigerwald:2021bro} DM have all been posited as potential catalysts for unusual SNe, including Ca-rich transients~\cite{Fedderke:2019jur}. The details of each DM model are mostly irrelevant to our analysis, but for concreteness, we will choose a PBH model with a single mass $M_{\rm PBH}$.

DM-induced WD ignition occurs when a DM transit deposits sufficient energy below the WD crust to induce a runaway fusion process and \sn explosion.
%The crucial effect needed for DM-induced WD ignition is sufficient energy deposition below the WD crust. This additional energy injection induces a runaway fusion process and \sn explosion. 
Refs.~\cite{Graham:2015apa,Graham:2018efk} show that a WD can undergo runaway thermonuclear fusion if a sufficient region is heated to a temperature above 1~MeV. The critical size, $\lambda_{\rm min}$, is set so that the fusion rate exceeds temperature diffusion, with numerical calculations~\cite{1992ApJ...396..649T} indicating:

\begin{align}
\label{eq:minlength}
   & \frac{\lambda_{\rm min}}{\text{cm}} \hspace{-0.05cm} =  \hspace{-0.05cm} \begin{cases}
  2.8 \times 10^{-5} \sqrt{\frac{5 \times 10^9 \frac{\text{g}}{\text{cm}^3}}{\rho_{\rm WD}}}, & \frac{\rho_{\rm WD}}{\text{g}/\text{cm}^3}  > 1.6 \times 10^{8}  \\[0.2cm] 
    10^{-4} \left( \frac{2 \times 10^8 \frac{\text{g}}{\text{cm}^3}}{\rho_{\rm WD}}\right)^{2}, &               \frac{\rho_{\rm WD}}{\text{g}/\text{cm}^3}  < 1.6 \times  10^{8} \,
\end{cases} 
\end{align}
%
%  2.8 \times 10^{-5} \left( \frac{5 \times 10^9 \frac{\text{g}}{\text{cm}^3}}{\rho_{\rm WD}}\right)^{1/2}, & \frac{\rho_{\rm WD}}{\text{g}/\text{cm}^3}  > 1.6 \times 10^{8}  \\
%    10^{-4} \left( \frac{2 \times 10^8 \frac{\text{g}}{\text{cm}^3}}{\rho_{\rm WD}}\right)^{2}, &               \frac{\rho_{\rm WD}}{\text{g}/\text{cm}^3}  < 1.6 \times  10^{8} \,.
    
The dynamical friction from a PBH transit heats up nuclei in a 
%For the ignition process, dynamical friction that is caused during the PBH transit can be estimated. 
%The accelerated, colliding nuclei heat up a 
cylinder of radius \mbox{$R \approx 10\, \sqrt{m_N/\Delta T} \,G_{\rm N} M_{\rm BH}/v_{\rm BH}$}, where \mbox{$M_{\rm BH}$} and \mbox{$v_{\rm BH} \sim v_{\rm esc} \sim 10^{-2}$} are the mass and speed of the PBH (of the order of the WD escape velocity $v_{\rm esc}$), $m_N$ is the nucleon mass, $\Delta T \sim 1\,\rm  MeV$ the temperature increase, and $G_{\rm N}$ is Newton's constant. 
Requiring that the heated radius is larger than $\lambda_{\rm min}$,
%the critical scale, needed to trigger runaway fusion $R > \lambda_{\rm min}$, 
we can calculate the minimum PBH mass capable of igniting a given WD by combining Eq.~(\ref{eq:minlength}), with a model for the WD density \cite{Timmes}: 

%we can calculate the minimum PBH mass that can ignite a WD of a certain mass per  Eq.~(\ref{eq:minlength}), which depends on the WD density, which can be estimated by a fit to a numerical model of Ref.~\cite{Timmes} 
%
\begin{align}
\label{eq:WDdensity}
    \rho_{\rm WD} = 5.8 \times 10^6 \, \frac{\text{g}}{\text{cm}^3} \left( \frac{1.41}{M_{\rm WD}} -1 \right)^{-1.87} \,. 
\end{align}

\begin{figure}[t!]
    \centering
    \includegraphics[trim=50 0 50 3,width=0.9\columnwidth]{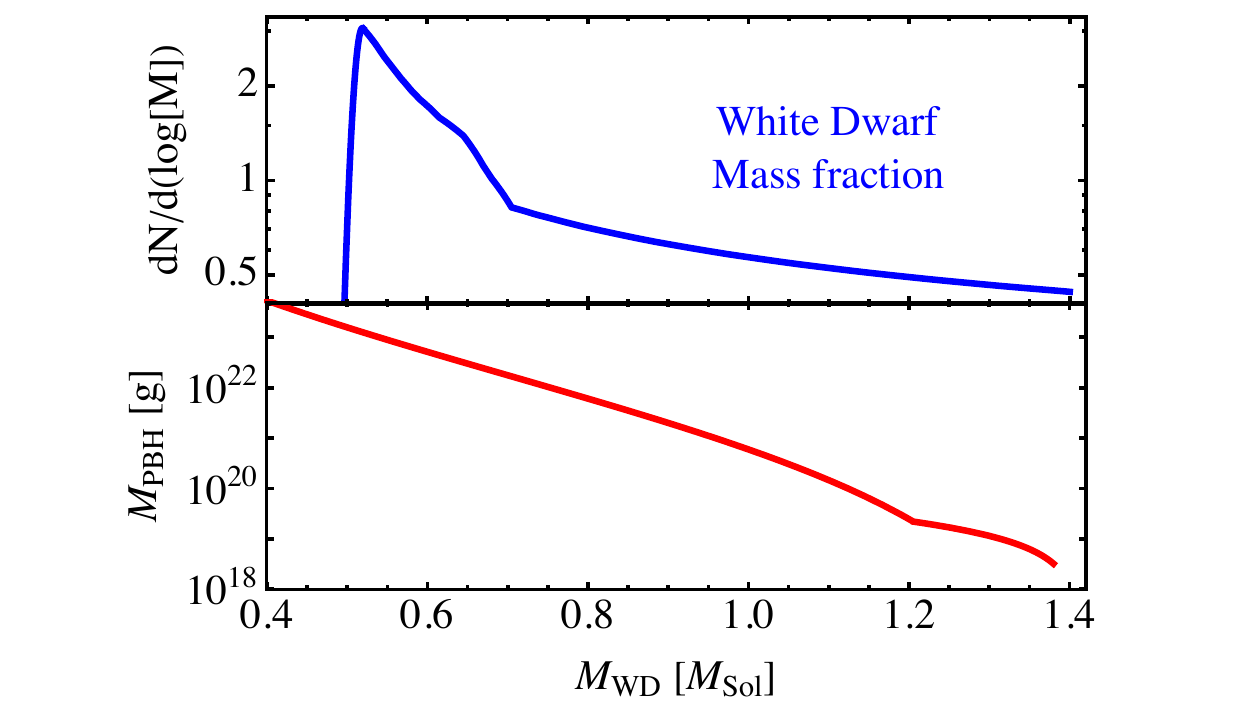}
    \vspace{1mm}
    \vspace{-0.2cm}
    \caption{Upper panel: The WD mass fraction as a function of the WD mass. Lower panel: The minimum PHB mass needed to trigger runaway fusion given a WD mass based on our analytic approximation. %\emc{It's a pity that the upper panel is cut off at 0.6 solar masses since this is still a relevant mass scale. I sort of like the usual mass function (i.e. not cumulative) since it shows more clearly that the mass peak is around 0.6.}
    }
 \vspace{-4mm}
    \label{fig:PBH}
\end{figure}

% total rate
Fig.~\ref{fig:PBH} shows the minimal PBH mass needed to trigger a WD explosion, together with the WD mass function. Because the WD density varies slowly in radius, even off-center transits have a high probability of producing ignition. Thus, we can express the WD explosion rate as:
%above which the probability, $f_{\rm ign}$, to trigger the WD explosion is of order one. Note that since the WD density is varying slowly, an off-center transit of the PBH will also lead to ignition. Thus, we can express the rate of explosion for a WD at a certain point as
%
\begin{align}
    \Gamma_{\rm ign} = \phi_{\rm DM} f_{\rm ign} = \pi R^2 \frac{\rho_{\rm DM}}{m_{ \rm DM}} v_0 \left( 1 + \frac{3}{2} \frac{v_{\rm esc}^2}{v_{\rm DM}^2}\right) f_{\rm ign} \,,
\end{align}
where $v_0 = \sqrt{8/(3 \pi)} v_{\rm DM}$, $v_{\rm DM}$ is the DM velocity dispersion at the considered position, $\phi_{\rm DM}$ the DM density, $R$ the WD radius, and $v_{\rm esc}$ the escape velocity from the WD. The factor $f_{\rm ign}$ encodes the probability of a PBH igniting the WD on transit. We will model this probability as a step function in the PBH mass, based on the simplistic parametrization discussed above. Note that a detailed simulation would be required to obtain exact values for  $f_{\rm ign}$, as discussed in Ref.~\cite{Montero-Camacho:2019jte}. In particular Ref.~\cite{Montero-Camacho:2019jte} shows that given the uncertainty on the ignition probability, currently no bounds on PBHs from WD observations can be made. A signal detection, however, would favor a particular ignition scenario.

To obtain the DM-induced ignition rate in a certain region we integrate the local rate over the WD distribution and obtain
\begin{align}
\label{eq:eventrate}
 \Gamma_{\Omega} = \int_{\Omega} dV \, \Gamma_{\rm ign} n_{\rm WD} \, ,
\end{align}
where $n_{\rm WD}$ is the WD number density and $\Omega$ is the region of interest. We will use this formalism to find the induced WD ignition rates in several cosmic environments. \\

%Here, I think go through a model, with the current Equation 1, and other important results. We specify asteroid-sized dark matter as a template system for this analysis, but also talk about the strange matter, phase transitions, etc. We probably also talk about the 0.6 Msun mass range for Ca-Gap Transients, and how this is weird from a stellar or binary evolution perspective -- but maybe this goes in its own section later in the paper.

%\newpage

\noindent {\bf \emph{Milky Way as a Template.}---} Most Ca-rich transients have been found near galaxies that have masses comparable to the Milky Way. This makes it reasonable to use the Milky Way as a ``template" to roughly estimate the distribution of Ca-rich transients around galaxies. We will define these analog galaxies as ``Milky Way equivalent galaxies" (MWEGs)~\cite{Bramante:2017ulk}.

Determining the radial distribution of Ca-rich transients in an MWEG requires models for the Milky Way dark matter density, stellar density, and star-formation history. For the DM density, we utilize a generalized Navarro-Frenk-White (gNFW) profile~\cite{1996ApJ...462..563N}, defined by:

\begin{equation}
\label{eqn:gNFW}
 \rho_\mathrm{DM}(r;\gamma,r_s,\rho_0)=\rho_0\left(\frac{R_0}{r}\right)^{\gamma}\left(\frac{r_s+R_0}{r_s+r}\right)^{3-\gamma}\,,
\end{equation}

\noindent where $\rho_0~=~0.42$~GeV~cm$^{-3}$~\cite{Read:2014qva} is the DM density at the solar position $R_0$~=~8.15~kpc~\cite{2019ApJ...885..131R}.
We consider a cuspy NFW profile with $r_s$~=~10.4~kpc and $\gamma$~=~1, and a cored profile with $r_s$~=~6.4~kpc and $\gamma$~=~0.01 representing values that are consistent with studies of the Milky Way DM density for different baryonic models~\cite{2019JCAP...09..046K}. For the dark matter dispersion velocity, which includes significant contributions from baryons in the inner regions where the dark matter density is highest, we utilize the fitting results of Ref.~\cite{Kafle:2014xfa}.

Constructing complete observations of the Milky Way WD population is difficult due to their low luminosity. Thus, we model the Milky Way WD population by examining the density distribution and star-formation history of their progenitors. We take the stellar density from the SDSS Milky Way tomography~\cite{SDSS:2005kst}, and calculate the star-formation history from the fitting function~\cite{Hopkins:2006bw}:

%calculate the white dwarf population throughout the galaxy, we model the historic star-formation rate following a fitting function~\cite{Hopkins:2006bw}

%Second, the stellar density is taken from a fit to \js{( or exponential disk )} observational data~\cite{Carlson:2016iis}, which covers the region close to the galactic center GC.  

%Finally, we need the fraction of white dwarfs, that have evolved over the lifetime of the MW. In order to find this value, we use the star formation rate, as fitted in Ref.~\cite{Hopkins:2006bw}, which is given by
%
\begin{align}
    \dot{\rho}_{\rm star} = \frac{h \left(a + b z \right)}{1 + \left(z/c \right)^d}\,,
\end{align}
where $h = 0.7$, and the coefficients have been fitted to \mbox{$a = 0.017$}, $b=0.13$, $c = 3.3$, and $d = 5.3$.

The redshift-time relation has been evaluated using 
%\tim{is this a good approximation? shouldn't we just use the real value, or does this produce too many problems given the need to do this fast and analytically? Is there a citation for this at least?}
%
%\begin{align}
%    \frac{\tau}{\text{Gyr}} \approx %\frac{28}{1 + (1 + z)^2}\,.
%\end{align}
%
\begin{align}
   & \tau(z)  = H_0^{-1} \int_z^\infty \frac{dx}{(1+x) \sqrt{\mathcal{E}(x)}}, \text{ with} \nonumber \\ & E(z) = \Omega_\Lambda + \Omega_{\rm M} (1+z)^3 + \Omega_{\rm R} (1+z)^4\,,
\end{align}
for which we use cosmological parameter values from~\cite{Planck:2018vyg}. For the stellar mass distribution, we use the Kroupa initial mass function~\cite{2003ApJ...598.1076K}, given by:
\begin{align}
   & \zeta_{\rm MW} = \frac{{\rm d}N_{\rm star}}{{\rm d}M} \\ \nonumber
 = &  \frac{\kappa}{M^\alpha}  \begin{cases}
    \kappa =0.07, \alpha = 2.3 ,& \text{if } \frac{M}{M_\odot} \geq 0.5 \\
     \kappa = 0.14, \alpha = 1.3,              & \text{if } 0.5 > \frac{M}{M_\odot} \geq 0.08 \\
     \kappa =1.78, \alpha = 0.3,              & \text{if } \frac{M}{M_\odot} < 0.08\,.
\end{cases} 
\end{align}

The WD mass upper limit of 8~$M_\odot$ stems from the lower-bound for electron-capture SN~\cite{1984ApJ...277..791N}, while the minimum WD mass depends on the stellar age, as:

%\tim{Say that Max comes from min SN mass, and min from the equation, will be easier to understand.} The mass bracket of stars of the age $\tau = \tau_{\rm max}- \tau_{\rm birth}$ that will have formed WDs is in the range 
%
\begin{align}
%\frac{M_{\rm max}}{M_\odot } = 8 \text{ and } 
\frac{M_{\rm min}}{M_\odot } = \min{\left[ \left( 1 - \frac{\tau_{\rm birth}}{\tau_{\rm max}} \right)^{-2/5}, \, 8  \right]} \, ,   
\end{align}
where $\tau_{\rm max} \approx \SI{10}{Gyr}$. Combined, we find a WD fraction:
\begin{align}
    f^{\rm MW}_{\rm WD} = \frac{\int_0^{\tau_{\rm max}} d\tau \dot{\rho}_{\rm star} \int_{M_{\min}(\tau)}^{M_{\rm max}} dM \zeta_{\rm MW}}{\int_0^{\tau_{\rm max}} d\tau \dot{\rho}_{\rm star} \int_{0}^{100 M_{\odot}} dM \zeta_{\rm MW}} \approx 0.035\,.
\end{align}
We have confirmed that this number does not strongly depend on the exact form of the stellar birth rate, and is a rather robust estimate for the WD fraction of the MW.

Assuming that every DM transit triggers a WD explosion, we obtain a total WD ignition rate of:

\begin{align}
\label{eq:NFWrate}
    R_{\rm MW} = (5 \pm 3) \times 10^{-4} \left( \frac{10^{24} \text{g}}{ M_{\rm PBH} }\right) \text{yr}^{-1} \,.
\end{align} 

Since this calculation includes significant uncertainties, it is best viewed as an order-of-magnitude estimate. \\

%Note that this number necessarily has substantial astrophysical uncertainties, and should be seen as a well-educated order of magnitude estimate. \\

\noindent {\bf \emph{Dwarf Model.}---} The Milky Way is surrounded by smaller structures known as dwarf spheroidal galaxies (dSphs)~\cite{2012AJ....144....4M}. While the total stellar population of dSphs pales in comparison to the Milky Way, they may dominate the DM/WD interaction rate due to a concordance of three factors.

First, dSphs contain a significant DM density, with mass-to-light ratios that normally exceed a factor of 100~\cite{2012AJ....144....4M}. Second, the velocity dispersion within dSphs is extremely low, often $\sim$10~km~s$^{-1}$. This significantly enhances the DM-WD cross-section due to gravitational focusing.

Third, the stellar populations of dSphs are old, increasing the fraction of stars that have evolved off of the main sequence and become WDs. Many dSphs, especially the subpopulation known as ultra-faint dSphs, show evidence for only a single star-formation episode near the epoch of reionization~\cite{2010ApJ...720.1225M, 2015ApJ...811L..18G}. While some larger dSphs, such as Carina, also have later, episodic, star-formation~\cite{2014A&A...572A..10D}, their star populations are still biased towards very old stars.

%Our MW is surrounded by a cloud of smaller structures, so-called dwarf spheroidal galaxies (DSG)~\cite{2012AJ....144....4M}. The DSGs are special in several regards. 
%First, the stellar populations are old, we assume all stars to be of order \SI{10}{Gyr} old. Second, the DM density is relatively large, and third, the DM velocity dispersion is significantly lower than in the MW. 
The confluence of those factors leads to a triggered SN rate in the dSph population that is similar to that expected from the main galaxy, despite significantly smaller star counts. To model the WD population in dSphs we use the initial mass function from Ref.~\cite{2018ApJ...863...38G}, which is given by:
\begin{align}
   \zeta_{\rm dSh} = 
   \frac{\kappa}{M^\alpha}  
 \begin{cases}
     \kappa = 1.0, \alpha = 1.3,              & \text{if } \frac{M}{M_\odot} \geq 0.08 \\
     \kappa =1.34, \alpha = 0.3,              & \text{if } \frac{M}{M_\odot} < 0.08\,.
\end{cases} 
\end{align}

We assume that dSphs have a uniform stellar age distribution at \SI{10}{Gyr}, which produces a larger WD fraction:
%Roughly assuming the stellar population to have a uniform age distribution at \SI{10}{Gyr} we find a larger WD fraction:
%
\begin{align}
    f_{\rm WD}^{\rm dSh} = \frac{ \int_{M_\odot}^{8 M_\odot} dM \zeta_{\rm dSh}}{ \int_{0}^{100 M_{\odot}} dM \zeta_{\rm dSh} } \approx 0.17\,.
\end{align}

We adopt a functional form for the stellar density profile from Ref.~\cite{Walker:2009zp}, and utilize the gNFW DM profile for both a cuspy profile ($\gamma$~=~1 with $r_c$~=~800~pc) and a cored profile ($\gamma$~=~0.1 with $r_c$~=~150~pc) (see, e.g.,~Ref.~\cite{Boldrini:2021aqk} for a discussion of the core-cusp problem). In dSphs, both the DM velocity dispersion and stellar kinematic profile are determined from kinematic data~\cite{Walker:2009zp}. Using these profiles, we calculate the triggered SN rate in a dSph with a given halo mass, half-light radius, and stellar mass function using Eqn.~\ref{eq:eventrate}. For example, given the input values for the dwarf Fornax from Ref.~\cite{2012AJ....144....4M}, we find an event rate of $\left(4 \pm 2 \right) \times 10^{-5}$ yr$^{-1}$ given a PBH mass of $10^{24}$ g, depending primarily on the DM profile. The expected total rate from all MWEG dSphs is about an order of magnitude larger. \\   

%
%\begin{figure}[t!]
%\vspace{+2mm}
%    \centering
%    \includegraphics[width=0.65\columnwidth]{figs/MWRrate.pdf}
%    \caption{Schematic of the DM-induced SN explosion rate in the MW and its dwarf spheroidal satellites.}
%    \vspace{-4mm}
%    \label{fig:schematic}
%\end{figure}

%\js{we probably won't show the 3D plot}
%In Fig.~\ref{fig:schematic} we show the position and expected signal strength for the MW with its known satellite galaxies, as listed in Ref.~\cite{2012AJ....144....4M}. The relative expected signal strength is color coded, with red being the highest and blue being the lowest expected event rates. We observe that the expected signal from the satellite sample is dominated by a few objects with the highest compactness.

\textit{Dwarf Distributions and Properties}--- As reported in Ref.~\cite{Grand:2021fpx}, state-of-the-art N-body simulations agree with the observed radial and size distribution of MW satellites. Thus, we can use the properties of the known MW satellites to extract statistical quantities that represent the MWEG dSph population, comparing our results to the expected triggered SN rate from the galaxy itself. %Those quantities will be used to compute the average expected event rate for DM-induced WD explosions in the satellites and compare it to the rate expected from the galaxy itself. 
The radial distribution of dSphs is well-described by a modified Gaussian distribution 

\begin{figure}[t!]
\vspace{-4mm}
    \centering
    \includegraphics[width=0.9\columnwidth]{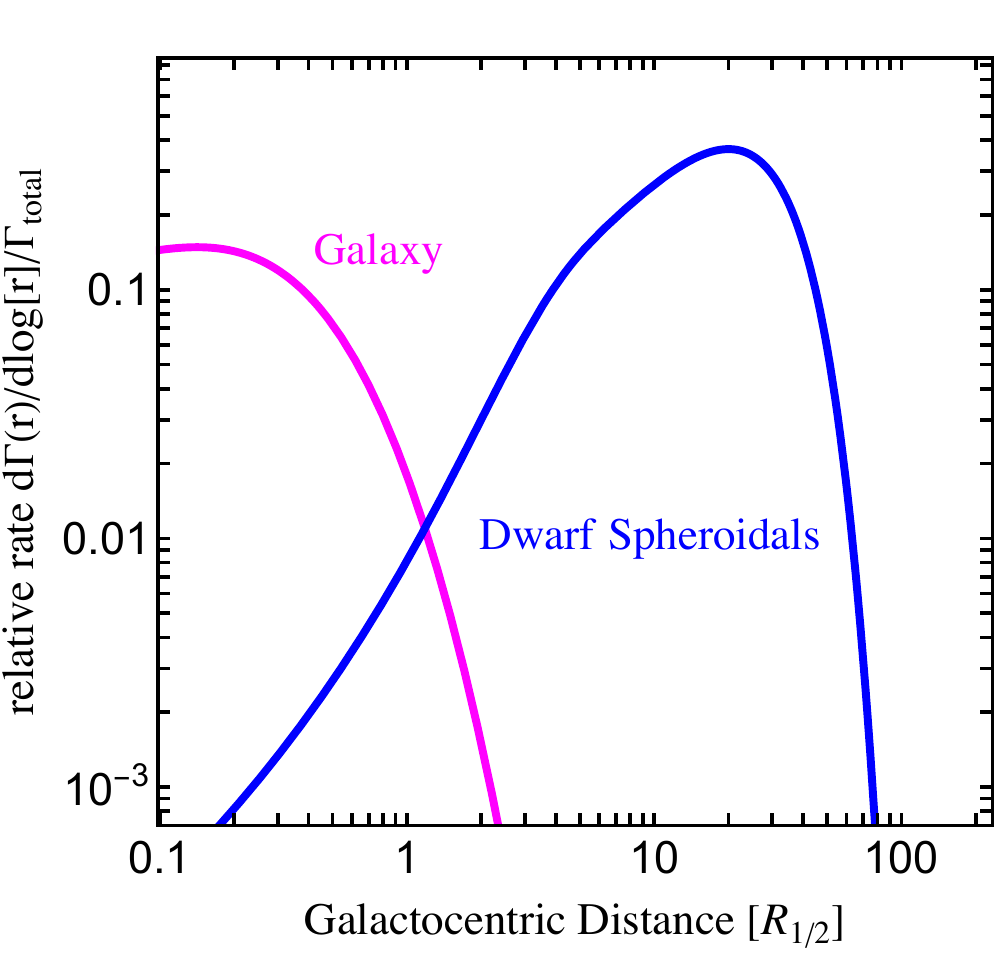}
    \caption{The modeled relative event rate per unit log-distance for triggered SN events stemming from the galaxy and dSphs as a function of the galactocentric distance. The best-fit SN rate from dSphs exceeds the Milky Way signal by a factor of $\sim$2--4.}
    \vspace{-4mm}
    \label{fig:radialnorm}
\end{figure}

\begin{align}
\label{eq:skewgauss}
    P^{\rm dSph}_r(\mu, \sigma, \alpha) = \left( 1 + \text{Erf}{\left(\frac{\alpha \left( r -\mu \right)}{\sqrt{2} \sigma }\right) } \right)\times \exp{\left[- \frac{\left(r-\mu \right)^2}{2 \sigma^2}\right] }
\end{align}
where $\alpha$ is a skewness parameter and 
$\text{Erf}$ is the Gaussian error function. The best-fit parameters, based on the MW dwarf sample in Ref.~\cite{2012AJ....144....4M} are $\mu = 12 \, \rm kpc$, $\sigma = 114 \, \rm kpc$, and $\alpha = 12$.  We generate a synthetic sample of dSphs populations around a galaxy, from which we estimate the total WD ignition rate expected from the dSph population: 

\begin{align}
 R_{\rm dSph} \approx 
 (12 \pm 8)\times 10^{-4} \left(\frac{ 10^{24} \text{g} }{M_{ \rm PBH}}\right) \text{ yr}^{-1}\,,
 %(2 - 4) R_{\rm MW} \, ,   
\label{eq:dSphrate}
\end{align}
which implies a ratio $R_{ \rm dSph}/R_{\rm MW}$ between 1 and 4. We discuss our synthetic dwarf population generation in App.~\ref{app:stats}. \\

%We first need to note that most of these Ca-Gap Transients (and indeed Type IA SN) are found surrounding Milky Way-like systems.

%Then we get into some numbers for the Milky Way in particular.

%This probably takes two columns, because there are going to be quite a number of equations here.
%

\begin{figure}[t!]
\vspace{-4mm}
    \centering
    \includegraphics[width=0.9\columnwidth]{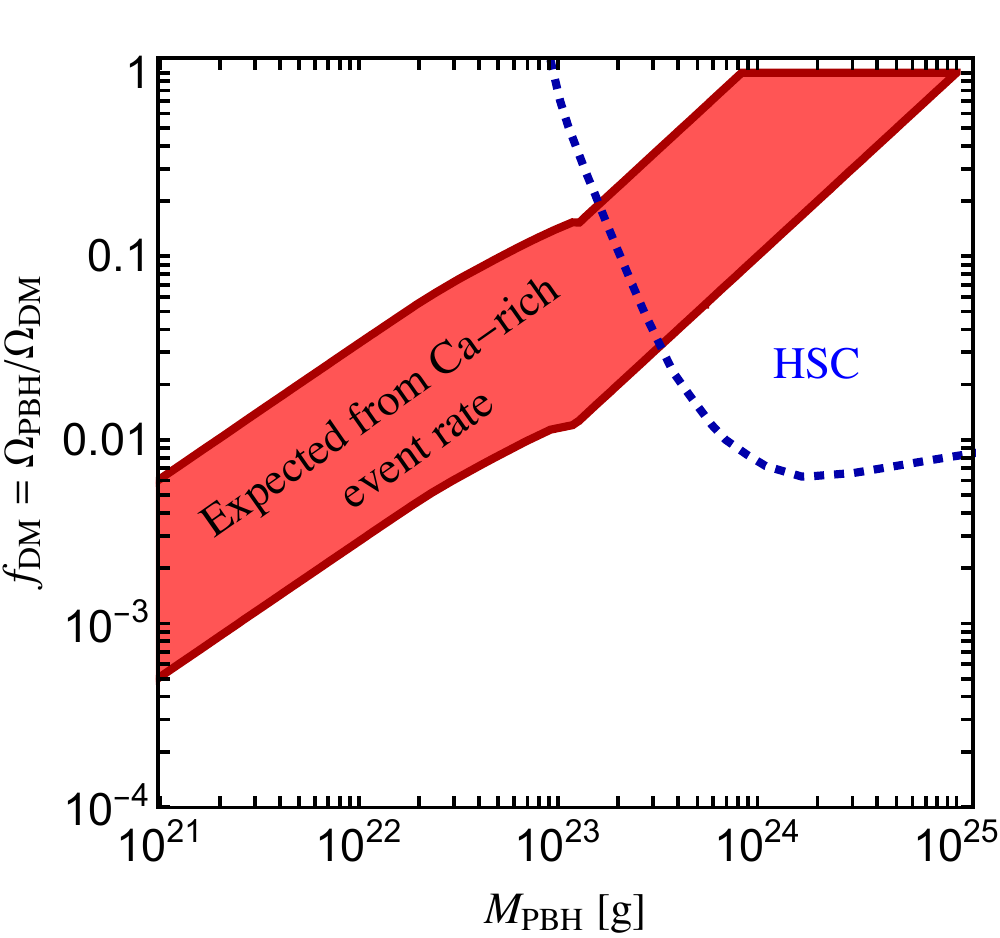}
    \caption{The combination of PBH mass and abundance (where $f_{\rm DM}$ represents the fraction of the DM mass composed of PBHs) required to produce a triggered SN rate compatible with Ca-rich Transient observations. The 95\% CL exclusion region from stellar microlensing in M31 by Subaru (HSC)~\cite{Niikura:2017zjd,Smyth:2019whb} is shown as a dashed blue line.
    %and the expected MW-SN event rate ($< 10^{-2}$ 1/yr) are shown~\cite{Graham:2018efk}.
    }
    \label{fig:parameterspace}
\end{figure}

\noindent {\bf \emph{Results.}---} %\tim{Do we need this paragraph at all? Can we just go straight to Figure 3?} With the above ingredients we can predict the spatial signal morphology expected for WD ignition that is triggered by a compact DM transit. 
%For the signal prediction from the MW, we integrate Eq.~(\ref{eq:eventrate}) in cylinder-shaped shells around the galactic origin with a radial thickness of one kpc. 
In Fig.~\ref{fig:radialnorm} we show the normalized, differential event rate in bins of width 1~kpc, under the assumption that $ R_{\rm dSph}  = 2  R_{\rm MW}$. The largest signal rate (by far) is expected in the galactic outskirts. In order to compare our results to observations, which include galaxies that are not exactly the same size or mass as the Milky Way, it is convenient to re-normalize the radial distribution of our model to the half-light radius of the galaxy.
%For the comparison to observations, however, it is convenient to normalize the radial distribution to the half-light radius of the system. 
We use $R_{1/2} \approx 6 \, \rm kpc$ as a benchmark value for the Milky Way to normalize our radial distributions. We overlay our theoretical expectations for the cumulative Ca-rich transient rate with the observations from Ref.~\cite{2017ApJ...836...60L}, see Fig.~\ref{fig:datacompare}. The data is best described with a relative dSph event rate of $ R_{\rm dSph}  \approx 3  R_{\rm MW}$, which is well within (and relatively conservative compared to)  our theoretical expectations. 

The total event rate of the Ca-rich Gap Transients was estimated to be \mbox{$R_{\rm obs}^{\rm total} = 3.19^{+ 1.45}_{-0.96} \times 10^{-6}$Mpc$^{-3}$yr$^{-1}$~\cite{2020ApJ...905...58D}.} This corresponds to a galactic event rate for an MWEG of \mbox{$R_{\rm obs}^{\rm Galaxy} = 4 \pm 2 \times 10^{-4} \text{ yr}^{-1}$.} We note that the event rates in Eqns.~\ref{eq:NFWrate}~and~\ref{eq:dSphrate} scales inversely with the PBH mass. Moving to smaller PBH masses increases the number of PBHs (and thus the total interaction rate), but also increases the minimum WD mass at which an interaction triggers an SN explosion. As can be seen from Fig.~\ref{fig:PBH}, transits of PBHs with masses of $10^{24}$~g can trigger SN explosions in WDs with masses as low as $0.4 \, M_{\odot}$. However, it is also possible that WD explosions in such low-mass starts will have a reduced detection probability compared to the detonation of more massive WDs~\cite{Graham:2015apa}. %Thus, PBH masses of $10^{24}~\text{g}$ should be considered as an upper bound, above which we do not get additional signal from the detonation of lower-mass WDs. 

%Given the scaling of our rate in Eq.(~\ref{eq:NFWrate}) 

%\tim{?? the PBH mass expected, assuming that the PBHs constitute all the DM in the universe, around $10^{24} \text{g}$.} As can be seen from Fig.~\ref{fig:PBH}, transits of PBHs of this mass will trigger WD explosions for WM masses above $M_{\rm WD} \sim 0.4 M_{\odot}$. However, it is possible that WD explosions of such low-mass stars will have a reduced detection probability in the visible spectrum~\cite{Graham:2015apa}. \tim{Thus, the mass of  $10^{24} \text{g}$ should be considered as an upper bound, as lower masses are expected as a result of reduced detection efficiency.}   

In Fig.~\ref{fig:parameterspace} we show the expected parameters in the PBH mass and fractional abundance plane. The PBH abundance is parameterized as a fraction of the total DM abundance needed to match observations~\cite{Planck:2018vyg}, $f_{DM} = \Omega_{\rm PBH}/\Omega_{\rm DM}$. We show an uncertainty band that encompasses the uncertainty of the rate computation and the theoretical uncertainty of our rate prediction.  This rate assumes a detection efficiency for Ca-rich Gap Transients calculated by Ref.~\cite{2017ApJ...836...60L}. In addition, we show the claimed 95\% CL exclusion region from a 7-hour-long dedicated search for microlensing of stars in the Andromeda galaxy, carried out using the Hyper Suprime-Cam (HSC) on the Subaru telescope, see Refs.~\cite{Niikura:2017zjd,Smyth:2019whb}. The upper limit is based on the detection of a single candidate event, whereas the team expected multiple detections if low-mass PBHs would dominate the DM abundance. Verification of the HSC limits would be highly desirable given our findings. Other constraints on PBHs in this mass range have been claimed in Refs.~\cite{Capela:2012jz,Defillon:2014wla}, but Ref.~\cite{Montero-Camacho:2019jte} showed that at the moment, the modeling and astrophysical uncertainties can not support the robustness of any of those bounds.

If the Ca-rich gap transient population includes additional, fainter, SNe than considered by Ref.~\cite{2017ApJ...836...60L}, the PBH mass range begins to be squeezed from both sides. High-mass PBH models become more strongly constrained by HSC observations, while low-mass PBH begin to be ruled out by the fact that such black holes could not detonate very low-mass WDs.  Thus, it is intriguing that the expected black hole mass and abundance range is in good agreement with the parameters needed to explain recent NANOGrav observations~\cite{DeLuca:2020agl}. \\

%We assume $100 \%$ detection efficiency, and 
%Note, that a systematically reduced detection efficiency would shift the expected parameter range to lower PBH mass values. 

%Taking into account this detection efficiency hypothesis, as well as the uncertainty on the Ca-rich event rate, and our expected uncertainty on the ratio of DSG events and galactic disc events $R_{\rm DSG}$, we obtain an expected mass window of %
%\begin{align}
    %M_{\rm PBH} =   10^{24} -  10^{25} %\, \text{g} \,, 
%\end{align}
%
%for the PBHs that would explain the observed Ca-rich event rate, given that they constitute all of the DM in the universe. Intriguingly, those PBH masses are allowed by other observational constraints~\cite{Graham:2018efk}.

\noindent {\bf \emph{Discussion and Conclusions.}---} In this letter we have argued that Ca-rich gap transient SN events may be explained by the ignition of relatively low-mass white dwarfs, triggered by the transition of heavy dark matter particles. In the minimal scenario that we focused on, these particles could be primordial black holes with masses of large asteroids, between $\sim 10^{21}~ \rm g$ and $\sim 10^{24}~ \rm g$, and radii of $\sim \mathcal{O}(1) \,n \rm m$ to $\sim \mathcal{O}(1) \, \mu \rm m$.

Two observations motivate this possibility. First, the Ca-rich transient events are inherently faint compared to standard SNIa and have peculiar spectral features that are difficult to model with standard astrophysical models. In the triggered ignition picture, this feature is expected, since the transits of sufficiently heavy ($>10^{21}$~g) dark matter can ignite even low-mass white dwarfs, with $M_{\rm WD} < 0.8  \, M_{\odot}$, which will produce a much fainter detonation signal. Intriguingly, this feature also significantly enhances the rate of Ca-rich transient events produced by DM interactions, due to the fact that the WD mass function peaks well below a solar mass.

Second, the spatial distribution of Ca-rich transients around a host galaxy differs significantly from the distribution of the SNIa events, see Fig.~\ref{fig:datacompare}, originating from binary systems involving at least one (heavier) white dwarf. While SNIa events match the stellar distribution of Milky Way-like galaxies, the Ca-rich transients do not. In the context of dark matter interactions, however, this behavior can be explained by the fact that dwarf spheroidal galaxies can dominate the event rate since they contain an old population of white dwarfs embedded in a compact dark matter halo with low-velocity dispersions.

%In addition the dark matter velocity dispersion in those regions is low, leading to a boosted rate, due to gravitational focusing. 

The possibility that dark matter is responsible for Ca-rich Gap transients is intriguing. This is particularly true since the simple scenario with primordial black holes with masses of order $\sim 10^{22} \rm g$ to $\sim 10^{24} \rm g$ is a unique region of parameter space, where the primordial black holes can make up a large fraction of the dark matter in our universe~\cite{Carr:2016drx,Carr:2020xqk}. It is a curious observation that this parameter range naturally explains the observed NANOGrav signal (see Ref.\cite{DeLuca:2020agl} for a detailed discussion of a scenario where a continuous PBH mass spectrum is shown, with a dominant peak in the mass range of $M_{\rm PBH} \sim 10^{21} \text{g} - 10^{23} \text{g}$). Note that also other scenarios with compact DM objects could explain the observed events. 

The expected mass range for the compact objects depends on the energy deposit efficiency. We note that recent 1D simulations by Ref.~\cite{Montero-Camacho:2019jte} have indicated that PBHs near the lower-bounds that we consider $\sim10^{21} - 10^{22}$~g may not be capable of exploding low mass $<$0.8~M$_\odot$ WDs, motivating a further exploration of the more massive PBH models consistent with our results. However, there are order-of-magnitude uncertainties in these limits and no full magnetohydrodynamic modeling has yet been performed. We leave a deeper investigation of such scenarios to future work.  \\

%\vspace{0.8cm}
%\vfill\null
%\columnbreak

\noindent {\bf \emph{Observational outlook.}---} The low luminosity of Ca-rich transients confines their discovery in ongoing transient surveys to a distance of $\sim$150 Mpc. The largest ongoing systematic survey is been carried out by the Zwicky Transient Facility, which classified 8 new events in its first 16 months of operation \cite{2020ApJ...905...58D}. The search volume will increase by more than two orders of magnitude once the deeper transient survey at LSST starts in 2024. However, the bottleneck for the identification of Ca-rich transients in the Rubin era will likely be dominated by spectroscopic resources. Hence, a more immediate route to explore the possibility that Ca-rich transients originate from WD interactions with DM could be to search for dwarf spheroidals at the locations of observed Ca-rich transients. For satellite galaxies brighter than \mbox{M$^{\rm dSh}_{\rm IR}\sim-7$ mag}, the average absolute magnitude of the known Milky-Way satellites \cite{2019ARA&A..57..375S},  JWST observations could have the sensitivity to resolve dwarf spheroidal galaxies for the current sample of Ca-rich transients from \cite{2020ApJ...905...58D}. Such an association would be a pivotal test for the proposed scenario.   

%\js{compare to re-scaled fornax}    

\section*{Acknowledgements}
AG acknowledges helpful communication with Dan Kasen and Josh Simon at the beginning of this project, and K.~De for help in accessing their data. We thank John Beacom, Valerio De Luca, and Rebecca Leane for helpful comments. JS and TL are, in part, supported by the  European Research Council under grant 742104. TL is also supported by the Swedish Research Council under contract 2019-05135 and the Swedish National Space Agency under contract 117/19.
AG acknowledges support from the Swedish Research Council under contract 2020-03444. EM acknowledges support from the Swedish Research Council under contract 2020-03384.

\bibliography{main}

%Figure Example
%\begin{figure}[tbp]
%\centering
%\includegraphics[width=.48\textwidth]{Pbg_plot_final.pdf}
%\caption{This is a Figure Caption}
%\label{fig:Pbg}
%\end{figure}

%\begin{multline}
%\label{eq:integral}
%I(E_\gamma) = \int_0^{z_{max}}dz \int_{L_{\gamma, min}}^{L_{\gamma, max}} \frac{dL_\gamma}{{\rm LG}(10)L_\gamma} \frac{d^2V}{d\Omega dz} \times \\
%\times \sum_X \Phi_{\gamma, X}(L_\gamma, z) \frac{dF_{\gamma, X}(L_\gamma, (1+z)E_\gamma, z)}{dE_\gamma}e^{-\tau(E_\gamma, z)}
%\end{multline}

%%

\newpage 
\onecolumngrid

\begin{center}
\textbf{\large White Dwarfs in Dwarf Spheroidal Galaxies:\\
A New Class of Compact-Dark-Matter Detectors}

\vspace{0.05in}
{ \it \large Supplementary Material}\\ 
\vspace{0.05in}
{Juri Smirnov, Ariel Goobar, Tim Linden and Edvard Mörtsell}
\end{center}
\onecolumngrid
\setcounter{equation}{0}
\setcounter{figure}{0}
\setcounter{section}{0}
\setcounter{table}{0}
\setcounter{page}{1}
\makeatletter
\renewcommand{\theequation}{S\arabic{equation}}
\renewcommand{\thefigure}{S\arabic{figure}}
\renewcommand{\thetable}{S\arabic{table}}

%\appendix

\section{Synthetic Dwarf Spheroidal Populations}
\label{app:stats}

It has been observed that the MW dwarf spheroidal galaxies appear to have a universal DM mass profile, well described by an NFW halo~\cite{Walker:2009zp}. Furthermore, correlations between core properties, such as the half-light radius $r_{h}$, the mass enclosed in $r_{\rm h}$, and the total stellar mass have been reported in~\cite{Walker:2009zp,2012AJ....144....4M}. Recent N-body simulations are in agreement with the universality observations, favoring the NFW halo hinted at by the kinematics~\cite{Grand:2022olu}. 
 As we are interested in the expected white dwarf ignition event rate from a typical set of dwarf spheroidal galaxies surrounding a host galaxy, we generate synthetic populations of dwarf spheroidals, following the prescription described below and determine the median expected event rate. 
 
A fit to the sample reported in~\cite{2012AJ....144....4M} reveals that the distribution of the log of the total stellar mass of dwarf spheroidals can be well fitted by a skewed Gaussian of the same form as in Eq.~\ref{eq:skewgauss}
\begin{align}
    \log_{10} \left[ \frac{M_{*}}{M_{\odot}}\right] = P^{\rm dSph}_{M}(\mu_{M}, \sigma_{M}, \alpha_{M})\,,
\end{align}
with $\mu_{M} = 2.9$, $\sigma = 2.4$, and the skewness parameter $\alpha = 5$. Generating the stellar mass $M_{*}$ from the above distribution, the correlations are implemented in the following way. The half-light radius is obtained from 
\begin{align}
     \log_{10} \left[ \frac{r_{h}}{\rm{pc}}\right] = \left( 0.86 \pm 0.2\right) + \left( 0.29 \pm 0.04 \right) \log_{10} \left[ \frac{M_{*}}{M_{\odot}}\right]\,,
\end{align}
where the errors are drawn from a Gaussian distribution. The total mass 
$M_{h}$ within $r_{h}$, which is also strongly correlated to the stellar mass, is generated by
\begin{align}
    \log_{10} \left[ \frac{M_{h}}{M_{\odot}}\right] = \left( 4.4 \pm 0.3 \right) + \left( 0.47 \pm 0.06 \right) \log_{10} \left[ \frac{M_{*}}{M_{\odot}}\right] \, ,
\end{align}
again with Gaussian errors, as indicated by the values above. This scaling behavior reproduces the correlations in the data set provided in Ref.~\cite{2012AJ....144....4M}, and the correlations are in agreement with the ones found earlier in Ref.~\cite{Walker:2009zp}. Two hypothesis are tested for the DM halo, one is a cuspy NFW, $\gamma =1$, profile with $r_0 = 800 \,\rm pc$ and two is a cored generalized NFW profile with $\gamma = 0.1$ and $r_0 = 150 \,\rm pc$, as discussed in Ref.~\cite{Walker:2009zp}. This variation in the central shape of the profile covers the uncertainty reported in Ref.~\cite{Boldrini:2021aqk}.  Finally, given this profile assumption, allows us to analytically calculate the profile of the expected velocity dispersion~\cite{Walker:2009zp}. 

Applying this method of generating synthetic dwarf spheroidal galaxies, we generate $10^4$ realizations of dwarf spheroidal galaxy sets, for several set sizes, varying the set size between $20$ and $40$, those set sizes correspond approximately to the number of dwarf spheroidal satellites of the Milky Way that have been discovered so far.
For each synthetic set, we compute the expected supernova ignition rate by DM transits. We find that the median of the even rate expected from the satellites of a galaxy is 
\begin{align}
   R_{\rm dSph} \approx (12 \pm 8)\times 10^{-4} \left(\frac{ 10^{24} \text{g} }{M_{ \rm PBH}}\right) \text{ yr}^{-1}\,. 
\end{align}
We observe that the expected event rate in the satellites is of the same order, or even somewhat larger than in the center of the host galaxy.  This is driven by three effects, one is that some satellites have relatively large DM concentration factors, then the DM velocities are significantly lower than in the host galaxy, which enhances the gravitational focusing effect, and finally, given that the satellite galaxies are old structures, their white dwarf fraction is larger than in the host galaxy.

%%A Figure using Include Graphics
%\begin{figure*}[tbp]
%\centering
%\includegraphics[width=.48\textwidth]{heatmap_alpha_sigma_negative_brightest.pdf}
%\includegraphics[width=.48\textwidth]{heatmap_alpha_sigma_negative_brightestwrong.pdf}

%\caption{A caption goes here. A caption goes here. A caption goes here. A caption goes here. A caption goes here. A caption goes here. A caption goes here. A caption goes here. A caption goes here. A caption goes here. A caption goes here. A caption goes here. A caption goes here. A caption goes here.}
%\label{fig:brightest50}
%\end{figure*}

\end{document}